\documentclass[
reprint,
 amsmath,amssymb,amsfonts,
 aps,
 pra,
longbibliography
]{revtex4-1}

\usepackage[normalem]{ulem}

\usepackage{graphicx}
\usepackage{dcolumn}
\usepackage{bm}
\usepackage{hyperref}
\usepackage{xcolor}

\begin{document}

\title{Low-power multi-mode fiber projector overcomes shallow neural networks classifiers}

\author{Daniele Ancora$^{1,3,5}$}
\email{daniele.ancora@uniroma1.it}
\email{daniele.ancora@cnr.it}
\author{Matteo Negri$^{1,3}$}%
\author{Antonio Gianfrate$^2$}%
\author{Dimitris Trypogeorgos$^2$}%
\author{Lorenzo Dominici$^2$}%
\author{Daniele Sanvitto$^2$}%
\author{Federico Ricci-Tersenghi$^{1,3,4}$}%
\author{Luca Leuzzi$^{3,1}$}%

\affiliation{$^1$Department of Physics, Università di Roma la Sapienza, Piazzale Aldo Moro 5, I-00185 Rome, Italy}%

\affiliation{$^2$Institute of Nanotechnology, Consiglio Nazionale delle Ricerche (CNR-NANOTEC), Via Monteroni, I-73100 Lecce, Italy}%

\affiliation{$^3$Institute of Nanotechnology, Soft and Living Matter Laboratory, Consiglio Nazionale delle Ricerche (CNR-NANOTEC), Piazzale Aldo Moro 5, I-00185 Rome, Italy}%

\affiliation{$^4$Istituto Nazionale di Fisica Nucleare, sezione di Roma1, Piazzale Aldo Moro 5, I-00185 Rome, Italy}%

\affiliation{$^5$ Epigenetics and Neurobiology Unit, European Molecular Biology Laboratory (EMBL Rome), Via Ramarini 32, 00015 Monterotondo, Italy}

\date{December 18, 2023}

\begin{abstract}
In the domain of disordered photonics, the characterization of optically opaque materials for light manipulation and imaging is a primary aim.
Among various complex devices, multi-mode optical fibers stand out as cost-effective and easy-to-handle tools, making them attractive for several tasks.
In this context, we cast these fibers into random hardware projectors, transforming an input dataset into a higher dimensional speckled image set.
The goal of our study is to demonstrate that using such randomized data for classification by training a single logistic regression layer improves accuracy compared to training on direct raw images.
Interestingly, we found that the classification accuracy achieved is higher than that obtained with the standard transmission matrix model, a widely accepted tool for describing light transmission through disordered devices. 
We conjecture that the reason for such improved performance could be due to the fact that the hardware classifier operates in a flatter region of the loss landscape when trained on fiber data, which aligns with the current theory of deep neural networks.
These findings suggest that the class of random projections operated by multi-mode fibers generalize better to previously unseen data, positioning them as promising tools for optically-assisted neural networks.
With this study, in fact, we want to contribute to advancing the knowledge and practical utilization of these versatile instruments, which may play a significant role in shaping the future of neuromorphic machine learning.

\begin{description}
\item[Keywords]
Multi-mode fibers, random projections, disordered photonics, neural networks, 

classification, MNIST.
\end{description}
\end{abstract}

\maketitle

\section{\label{sec:Intro}Introduction}

A sound understanding of the enormous success of Neural Networks (NNs) in learning processes and inference tasks is still lacking. 
The fundamental point is to understand why such architectures, which can have even billions of parameters, do not severely overfit data, as predicted by statistical learning theory and the so-called \emph{bias-variance tradeoff} (see for example \cite{hastie2015statistical} or \cite{mackay2003information}).
The abundance of learnable parameters, in fact, is arguably the most universal feature in the zoo of NN architectures.
Interestingly, it is known that, given a chosen NN architecture, most of the model parameters adapt little-to-nothing during the learning procedure \cite{chizat2019lazy,geiger2020disentangling}, suggesting that random projections may play an equally important role in NNs.
Recent works, in fact,  have shown that it is possible to train a simple two-layer model by learning only the upper layer, interpreting the first one as a random projection \cite{gerace2020generalisation,goldt2022gaussian}.
These results were strengthened further by Baldassi et al. \cite{baldassi2021learning}, who proved that increasing the dimension of the random projection leads to the production of wide and flat regions in the loss landscape (the function that is minimized during the training of the model), which are related to the good generalization properties in neural networks. \textcolor{black}{The ability of a neural network, that has been trained over a given dataset ({\em training dataset}), to  generalize well is the ability of displaying good performances when applied to data over which it was not trained ({\em test dataset}). 
In the framework of the loss landscape description an improvement in the generalization means that models that lie in flat regions make less mistakes when they classify previously unseen data.}
Finally, recent evidence is provided \cite{perugini2022} that the way the random projection is chosen is fundamental to determining the generalization properties of the upper layer of these simple models.
This suggests that different classes of random (possibly non-linear) projections impact differently on the performance of the models. 

In this context, we are interested in studying hardware random projectors, such as those employed in the field of photonic neuromorphic computing \cite{wetzstein2020inference,de2019photonic}.
The advantage of using optical neural networks (ONN) is that neurons can interact by exploiting light scattering \cite{lin2018all,zhou2021large, li2021spectrally} and photon interference \cite{abu1987optical,shen2017deep} at the speed of light.
Tools for shaping and controlling the light-field \cite{dong2019optical} are becoming so versatile that the field is under constant development, aiming at high-speed, high-throughput optical-based computing architectures.
All-optical neural networks \cite{lin2018all, luo2019design}, in particular, have the potential to be great tools for fast computation, 
though they often
require an accurate modeling of the optical system to perform consistent back-propagation update \cite{Spall:22}.
However, the fine-tuning of the optical parameters is challenging due to discrepancies between the response of the real system and the physical model employed to describe the architecture.
This \emph{reality gap} often reduces the expected performance of the network \cite{zuo2019all,wright2022deep}, requiring additional corrections at software-level \cite{zhou2021large}, training enforcement via hybrid strategies \cite{Spall:22}, or employing NNs to more accurately model the optical response of the system \cite{wright2022deep}.

In this rapidly evolving scenario, the class of random projections realized by multi-mode fibers (MMF) are promising candidates for realizing ONNs.
These devices scramble the photons due to scattering events occurring during the light-field propagation, yielding to the formation of speckle patterns that are, in fact, random projections.
Although the light transmission can be regarded as a linear process \cite{popoff2010measuring} in which input modes are coupled with output modes via a complex transmission rule, interference takes place when dealing with the measurement of the light-field intensity.
Since the detection is nonlinear, MMFs can be used \cite{tanaka2019recent}  to classify time-domain waveforms (using saturation effects as further nonlinearity) \cite{paudel2020classification}, in pattern classification of 2-bits sequences \cite{porte2021complete}, or for binary (human, not human) facial recognition \cite{takagi2017object}.
Furthermore, when dealing with more complex classification tasks, high-power laser pulses were employed to trigger the nonlinear response of the fiber itself \cite{teugin2021scalable}.
Due to the increasing interest in the employment of MMFs as random projector computing devices, we decided to study their behavior in carrying out classification tasks in a linear, low-power continuum regime. 
Although our MMF-based optical neural network does not employ feedback, we will see how its performances in classification are considerable, as in reservoir computing systems \cite{saade2016random,du2017reservoir,ballarini2020polaritonic,chen2020classification}.

We do this by comparing the performance of the physical neural network to that obtained with random Gaussian linear projections and to that of a transmission matrix approach, the model commonly used to describe
light propagation in disordered structures \cite{popoff2010measuring,ancora2022transmission}.
We perform our study statistically, shuffling the training set to assess the average behavior of the 
optical computing
under different training and initialization conditions.
Remarkably, a single MMF provides simultaneously two independent (though deterministically linked) projections at both edges of the fiber, which we study separately using different saturation regimes.
Here, 
we show that the real physical MMF leads to accuracy higher than its corresponding transmission matrix model, highlighting the \textit{reality gap} 
between model theory and experimental results.
To assess the reason of this performance gap, we study the characteristic of complex-valued random projections in terms of flatness of the local energy landscape, proving that the MMF projection is more robust than those provided by alternative datasets.
Additionally, we characterize the behavior of a hardware-based neural network using optical fibers in terms of the numbers of the modes employed.
We have set up our study not for achieving the best performance in classification tasks, but rather to deepen the understanding of physical neural networks against their physical model, giving insights on the usage of MMF fibers for optical computation.

\section{Materials and Methods}
In a low-power regime, a generic multi-mode fiber transports the electromagnetic field via a linear process \cite{popoff2010measuring} so that the light propagation can be described using a simple multiplication of the input signal by a matrix that encodes the transmission rule:
    \begin{equation}
    \label{eq:linearT}
        \textbf{y}  = \mathbb{T} \textbf{x}.
    \end{equation}
In this descriptive model, $\textbf{x}$ is the controlled input, $\mathbb{T}$ is the (complex-valued and typically unknown) transmission matrix of the medium, and $\textbf{y}$ is the output field.
Despite its propagation, the way we measure the MMF output is not linear for two reasons.
First, photons carry complex signals, i.e., the electromagnetic field associated to each propagation mode is characterized by amplitude and phase.
Current electronic devices cannot follow the rapid oscillation of the field, which makes impossible the measure of the phase information. 
Assuming  the possibility that the readout  is also perturbed by an additive noise $\varepsilon$, the camera only sees the noise affected intensity distribution:
\begin{equation}\label{eq:detection}
    \left| \textbf{y} \right|^2 =\left| \mathbb{T} \textbf{x}\right|^2 + \varepsilon.
\end{equation}
Second, the camera has a well defined sensitivity range that depends on each pixels capability to store intensity change. 
If the signal reaching a given pixel exceeds the sensitivity, the measure gets clipped at the peak (overexposure) or at the bottom (underexposure). 
In analogy to machine learning terminology, the measuring process can be described by a non-linear \emph{activation function} $\sigma(\cdot)$ that acts on the result of a complex-valued linear transmission, $\mathbb{T} \textbf{x}$.
For instance, the camera recording process can be represented using the saturating linear transfer function (SatLin):
\begin{equation}
\label{eq:MMFsimulation}
    \sigma\left( \mathbb{T} \textbf{x}\right) = \min \left( \max  \left(d,\left| \mathbb{T} \textbf{x} \right|^2  + \varepsilon \right), 2^b-1 \right),
\end{equation}
where the quantity $d$ is the intensity threshold under which the measure is not recorded, and $b$ is the bit depth of the camera.

\begin{figure*}
    \centering
    \includegraphics[width=0.95\linewidth]{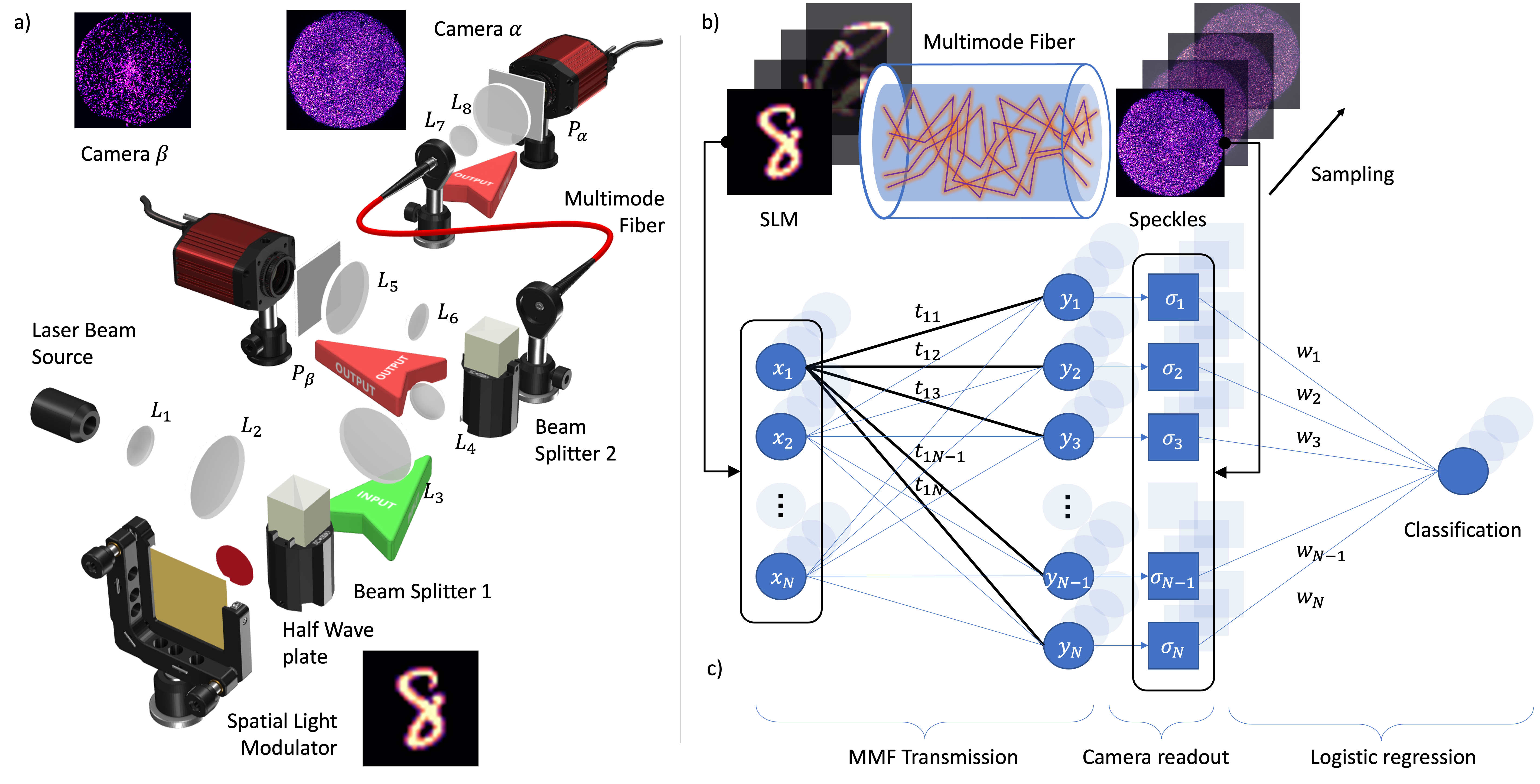}
    \caption{Schematics of the shallow optical neural network with the MMF. Panel a), simplified scheme of light transport through multi-mode fibers.
    The MNIST data set, modulated by a spatial light modulator, enters the MMF on the input facet. 
    During propagation, the light gets scrambled with a random but deterministic process, giving rise to the speckle pattern measured in the camera.
    Panel b), corresponding neural network interpretation of the light propagation scheme. 
    The MNIST set constitutes the input vector of a linear complex layer with static weights. 
    The non-linear operation is determined by the camera that reads the intensity of a complex field.
    Successively, a linear classification layer is trained using the output of the fiber.
    Panel c), scheme of the imaging setup.
    }
    \label{fig:scheme}
\end{figure*}

These considerations make the readout of a coherent field non-linear, as well as its inverse transmission recovery problem
\cite{van2010information, popoff2010measuring,Popoff10b,Aulbach11,Popoff14,Rotter17,ancora2022transmission}.
Such matrix can be estimated by using the four-phases method \cite{popoff2010measuring}, Bayesian optimization \cite{dremeau2015reference}, or iterative Gerchberg-Saxton schemes \cite{huang2021generalizing, ancora2022speckle}.
However, the characterization of the device in terms of its transmission rule is not the main scope of this paper, nor to circumvent the limitations of the measuring process.
Instead, we want to study the multi-modal random-projection nature of the fiber to perform optical computing.
In the neural network framework, the fiber can be seen as an optical analogous of a densely-connected  network composed by a single ``hidden layer" with fixed weights \footnote{The underlined network is one-layer deep because it can be described by a single matrix multiplication as in eq.\ref{eq:linearT}.}.
In this shallow architecture, the MMF layer already contains a particular realization of static weights (the transmission matrix $\mathbb{T}$), which depends upon the physical status of the optical fiber.
This property allows to perform random -but deterministic- projections at the speed of light using a fixed transmission rule, which can be read out by the camera.
Given these considerations, the MMF is a good candidate to perform non-linear optical computation using continuous laser source even using inexpensive and large (thus easier to handle in a setup) optical multi-mode fibers.
In particular, if we let just a few modes propagate into the input facet of an MMF that supports many more, all the output modes will be activated, implying a mapping of the kind few-to-many.
In this latter case, the optical hidden layer (i.e.,  MMF {\em and} camera) can perform densely-connected random projections on a higher dimensional space.

The goal of this study is to carry out image classification by concatenating a software-trained linear layer to the measured output of a MMF, produced by inserting a given image from the dataset into the input edge of the fiber.
We choose to approach the MNIST classification problem in order to carry out a widely studied non-linear task.
The only parameters that we train are those of a simple logistic regression layer, which is known to achieve poor performances on the standard MNIST set, reaching a maximum classification accuracy of 92.7\% \cite{ballarini2020polaritonic}.
Exploiting random projection provided by the MMF, an optical device that is known to be linear, we compare with the performances obtained using reference datasets.
In this study, we train the parameters of the logistic classifier using six different input datasets:
\begin{enumerate}
    \item \textit{Original MNIST.} The standard MNIST dataset, constituted by images of $l \times l$ pixels. The accuracy performance of this set is the baseline of our study.
    \item \textit{Upscaled MNIST.} Each image at the original resolution is expanded by a factor $L/l$ using a linear spline interpolation to reach the target size of $L\times L$.
    \item \textit{Randomized MNIST.} The MNIST set is linearly multiplied with a Gaussian random matrix with positive entries. This maps the dataset into a higher-dimension space, producing images with a side $L \gg l$ pixels.
    \item \textit{MMF $\alpha$-cam.} The speckled output of the MMF is recorded with a resolution of $L\times L$ pixels. Each speckle pattern is the result of sending a MNIST image on the input edge of the fiber, recording the output after disordered propagation. The patterns in input are intensity-modulated in real space, and have size of $l\times l$. 
    \item \textit{MMF $\beta$-cam.} Same as the previous one, with the speckles being recorded on the same input facet as that of the light injection. A relatively small portion of the light propagating forward is internally reflected and comes back towards the input edge. This determines a different speckle realization that we acquire as an independent measurement.
    \item \textit{MMF $\alpha$-simulated.} The transmission is characterized retrieving its corresponding matrix $\mathbb{T}$ using the SmoothGS protocol \cite{ancora2022speckle}. The inferred transmission is used to simulate the propagation of the MNIST set using Eq. (\ref{eq:linearT}), recording the simulated speckle pattern by storing only the squared modulus.
\end{enumerate}
All the datasets were used for a supervised training, in which the image of the MNIST set is associated to the number that represents, and the speckle image is associated to the classified number corresponding to the MNIST image impinged onto the fiber. 
Further details of the training procedure can be found in the Appendix section \ref{app:nn_details}.
To isolate any possible dependence on the problem size, we choose to set the size of the randomized and up-scaled MNIST sets to have the same dimension as the recorded fiber output.
This implies that the same number of parameters are trained while solving the classification problem for every dataset, the only exception being the original set.

\section{Results and Discussions}
\label{sec:Results}
In the following, we report the average results obtained by running independent logistic regressions on each dataset, comparing the classification accuracy on a test-set composed by $1000$ numbers isolated from the original one.

\begin{figure}
    \centering
    \includegraphics[width=1.\linewidth]{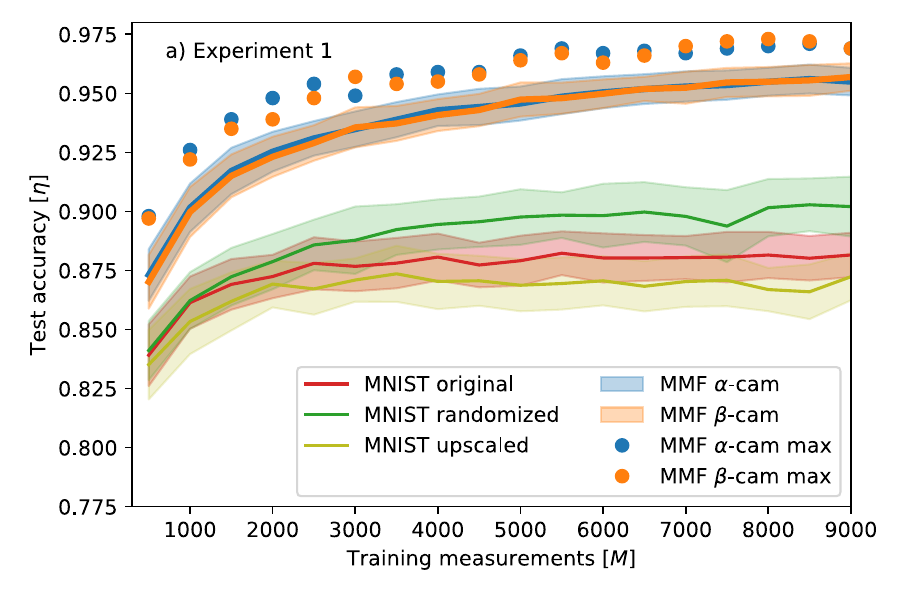}
    \includegraphics[width=1.\linewidth]{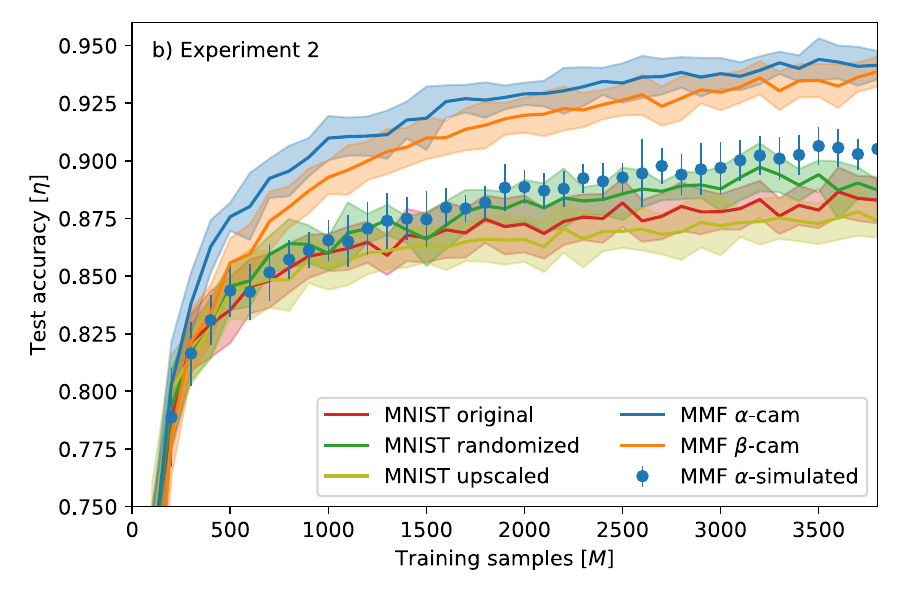}
    \caption{
    Logistic regression performance using different training datasets.
    a) Experiment 1, we train the model with up to nine thousand images. 
    The original MNIST set ($l=28$) was used as reference performance (red line) for the logistic regression, together with its upscaled (light green) and randomized (green) versions.
    Training the classificator with MMF-transformed speckle images (blue and orange curves) exhibited the highest accuracy.
    We sampled $100$ independent trainings, and we reported with the asterisk plot the best performance obtained with the fiber.
    b) Experiment 2, we train using up to 3800 images. Compared to panel a), we include also the output of the simulated fiber.
    The simulation was done by recovering the transmission matrix of the optical element, and using a complex linear transmission model.
    The simulated model performs better than the MINST set but did not reach the same performance as its experimental counterpart.
    }
    \label{fig:accuracy}
\end{figure}

\paragraph*{Performance of different class of projectors.}
In experiment 1, we used $10^4$ MNIST images, randomly picking up to $9000$ images for training and $1000$ images for testing, using $L=600$.
We repeat the parameter optimization for a total of $T=100$ times,  varying the number of training samples for statistical purposes.
\textcolor{black}{To test the robustness of our results after training, we compute the test accuracy, which is the fraction of correctly classified data points in the test set.}
From Fig.~\ref{fig:accuracy}a, we see how the performances of the MNIST dataset (original, randomized, and upscaled) are similar one to another.
The classification problem, in fact, is well known to be a non-linear task, and hardly generalizes using a linear model alone.
Instead, using the MMF higher performances are achieved, approaching 96\% test accuracy on average on the largest set used (9000 train examples).
We stress that this accuracy is not high in absolute terms since deep neural networks with convolutional layers have been able to reach more than 99\% test accuracy on MNIST 
\cite{lecun1998gradient}, with modern deep architectures raising even up to 99.91\% \cite{an2020ensemble}. 
However, we are interested in the study of the most simple ONN architecture, consisting only of a hardware random-projector layer followed by a linear classifier.
With this straightforward setup, the MMF permits to substantially improve the results obtained against a plain linear classifier (88\% accuracy with 9000 train examples). 

We point out that we did not use the entire MNIST dataset (composed of 60000 images for training and 10000 for testing) but a fraction of it; by increasing the number of training samples, the plot trend in Fig. \ref{fig:accuracy} suggests that there is room for further improvement.
Already after $\sim500$ samplings, the gain provided by the 
ONN
approach starts to become evident, and with only 9000 images, we can achieve performance hitting $\sim97 \%$.
To achieve the highest accuracy with the experimental data (blue and orange dots in the plot of Fig. \ref{fig:accuracy}a), we tested a $100$ independently initialized optimizations.
Interestingly, the performance are independent of the microscopic MMF arrangement, as the two different transmission rules determined by the $\alpha$ and $\beta$ detections perform identically.
As a final note, we decided not to tune the hyperparameters of the classifier, so we can expect that their meticulous choice (mainly the $l2$-regularization strength and the stopping threshold) could improve the accuracy curves for all the datasets.
In fact, we are not interested in the absolute numbers: our scope is to highlight the improvement determined by the physics of interactions of the MMFs, and the performance gain provided by the fine tuning of the hyperparameters with respect to each dataset would not change the main message of our work.

In experiment 2, we take a different static configuration of the fiber (i.e., characterized by another realization of  $\mathbb{T}$) that we probe with an alternated sequence of random and MNIST images. 
Differently from experiment 1, here we use the random patterns in input (and the related projection) to characterize the transmission matrix of the fiber using the SmoothGS protocol \cite{ancora2022speckle}.
We do this so that we can use the inferred $\mathbb{T}$ to simulate the propagation of the MNIST dataset through the fiber, obeying Eq. \ref{eq:detection}, and compare the classification performance of the linear model against that of the actual experimental measurements.
To make a fair comparison with simulated data, we tune the exposure time of the $\alpha$-cam to avoid saturated measurements.
Interestingly, we found out that training the logistic regression with the $\alpha$-simulated speckles does not perform well like the measured data.
The accuracy achieved is better than the direct MNIST set but worse than what was obtained using the experimental speckles (Fig.~\ref{fig:accuracy}b).
We observe, then, a "reality gap" that may be due to the presence of noise and other experimental non-linearities, which are not included in the way we model the physics of the system of Eq. \ref{eq:MMFsimulation} at low power.
It may be conjectured that non-linearities, also studied in the framework of computational optics with much more intense pulsed light \cite{teugin2021scalable, oguz2022programming}, already contribute at the lower intensities that we have been using in our experiments.
Compared to the setup used in \cite{teugin2021scalable}, we employed an energy density that is almost three orders of magnitude lower, also determined by the fact that we employed MMFs with large cores of $1mm$.
On the other hand, the data acquired by the $\beta$-cam was intentionally strongly underexposed (see the App. \ref{app:underexposure}).
By doing so, we notice that considerable thresholding has only a marginally negative impact on the performance. 
Even when the camera loses most of its signal, the accuracy of the classifier drops only by a factor of $\sim 2\%$, if compared with a better filling of the camera dynamic range in Fig.~\ref{fig:accuracy}b.
This little performance-drop enforces the idea that the MMF provides a class of random transformations that are particularly robust in carrying out classification tasks.


\begin{figure}
    \centering
    \includegraphics[width=1.\linewidth]{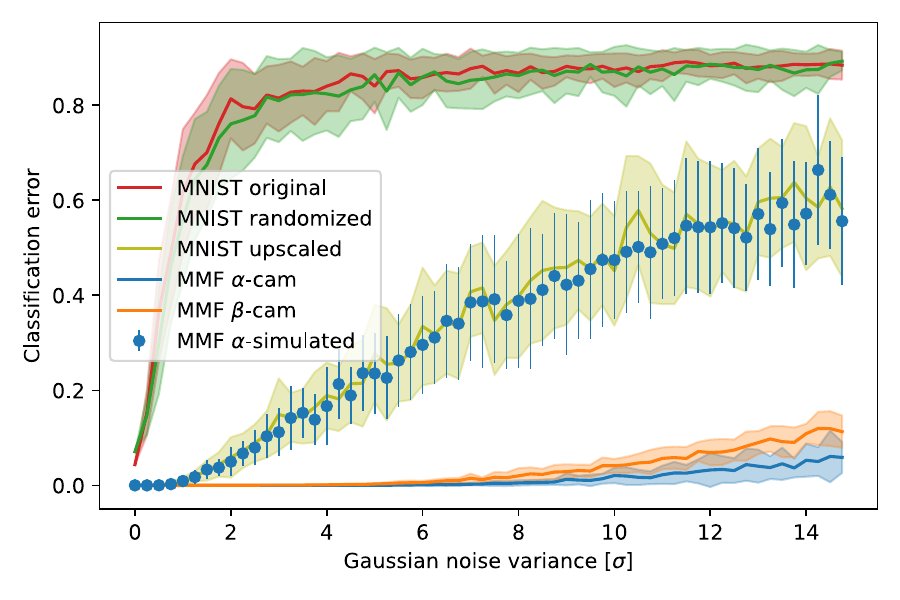}
    \caption{Local energy profiles for models trained on the different projected datasets. Each point corresponds to the train error of configurations sampled with multiplicative noise around the reference, averaged over 30 samples. The reference configurations are models trained on 3800 examples. Error bars show the standard deviation of the error distribution.}
    \label{fig:local_energy}
\end{figure}

\paragraph*{Accuracy of random projections and behavior of the training error.}
With this study we have set a testing ground for different random projectors used to pre-train the MNIST dataset, looking for those enforcing classification. 
In order to understand why the best accuracy results are obtained with MMFs, we study a measure of the flatness of the energy landscape (i.e., the train error) around the different model solutions. 
\textcolor{black}{Flatness is supposed to correlate well with generalization properties \cite{baldassi2021learning,jiang2019fantastic,locentfirst,unreasonable,entropysgd,pittorino2020entropic}, meaning that it can provide insights into how the geometry of the projected space influences the classification errors of new data points.}
We use the method of the \emph{local energy} to measure the flatness (see  \cite{baldassi2021learning} and references therein), that consists in adding a multiplicative Gaussian noise to the model parameters, sampling configurations with a given noise, and eventually computing the average fraction of misclassified data points (see Appendix section \ref{app:local_energy_appendix} for all the details). 
Performing this procedure for increasing noise values yields an estimate of the flatness of the reference configuration.
In Fig. \ref{fig:local_energy}, we see that the local energy profile correlates well with test accuracy shown in Fig.~\ref{fig:accuracy}: the flattest the solutions, the better the test accuracy. { The only exception to this is the upscaled dataset, which has the same local energy profile as the simulated dataset but shows a lower test accuracy (we discuss this point in section~\ref{sec:Perspectives}).}
A remarkable feature of the local energy profiles of MMF solutions is that they appear stable up to the noise of the order of $10$ times the signal-to-noise ratio. 
This robustness to noise might be the reason for the excellent generalization performance \textcolor{black}{on previously unseen data.}
This evidence supports the idea that MMFs are promising candidates for optical neural network computing.
The models trained on MMF-projected data show very low local energy variation. 
On one hand, this confirms the current idea in the literature that flatness correlates with generalization; and, on the other hand, it raises the question of why MMFs exhibit such a conceptual difference with their idealized model.
This reality gap could signal the presence of something not yet taken into account in the theoretical description of the physics of experimental set-ups with MMFs used in low-power mode.

\begin{figure*}
    \centering
    \includegraphics[width=1.\linewidth]{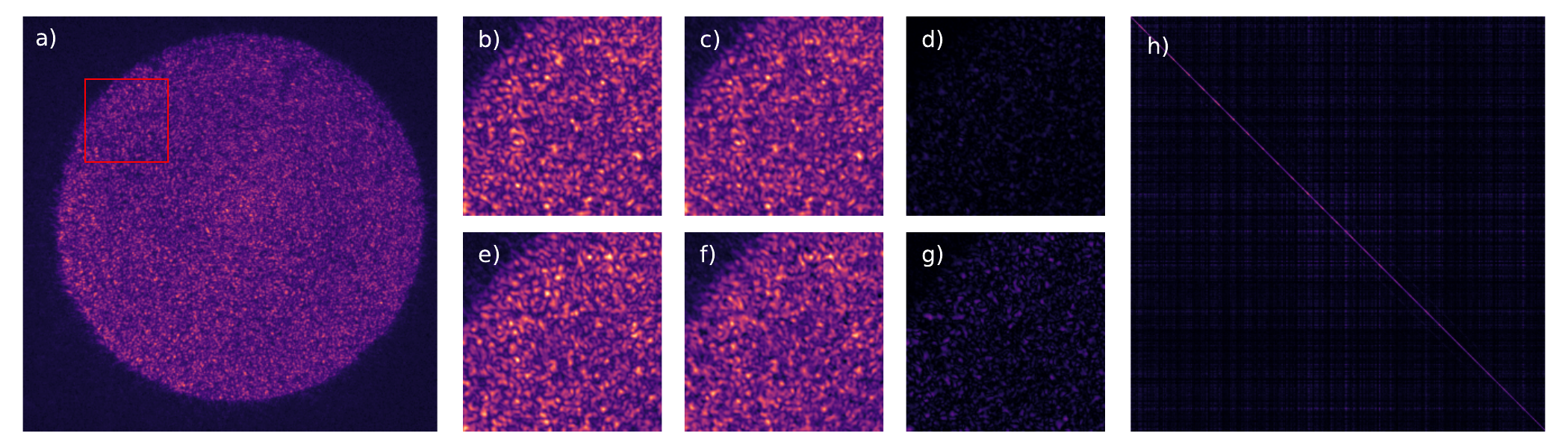}    
    \caption{
    a) MMF $\alpha$-camera speckle output after fiber propagation in experiment 2. The red box highlight a sub-region magnified in panel b) taken from the training dataset. 
    c) simulated speckle output after transmission matrix recovery, and 
    d) absolute difference between real and simulated speckle pattern.
    \textcolor{black}{
    Using Eq.\ref{eq:overlap}, we can compare the average similarity of the measured and simulated speckles of all seen random modes, obtaining $\rho_{train} = 0.865 \pm 0.082$.
    }
    e) speckle output recorded from the test set (not used for training),
    f) corresponding simulated output using the recovered transmission, and
    g) absolute difference between real and simulated data.
    \textcolor{black}{
    Similarly, the average similarity of all unseen random modes is  $\rho_{test} = 0.775 \pm 0.059$.
    }    
    h) focusing operator calculated using the recovered transmission of the output channels involved in the formation of the speckle in the red-box.
    }
    \label{fig:speckleout}
\end{figure*}

\begin{figure}
    \centering
    \includegraphics[trim={20 20 20 20},width=1.\linewidth]{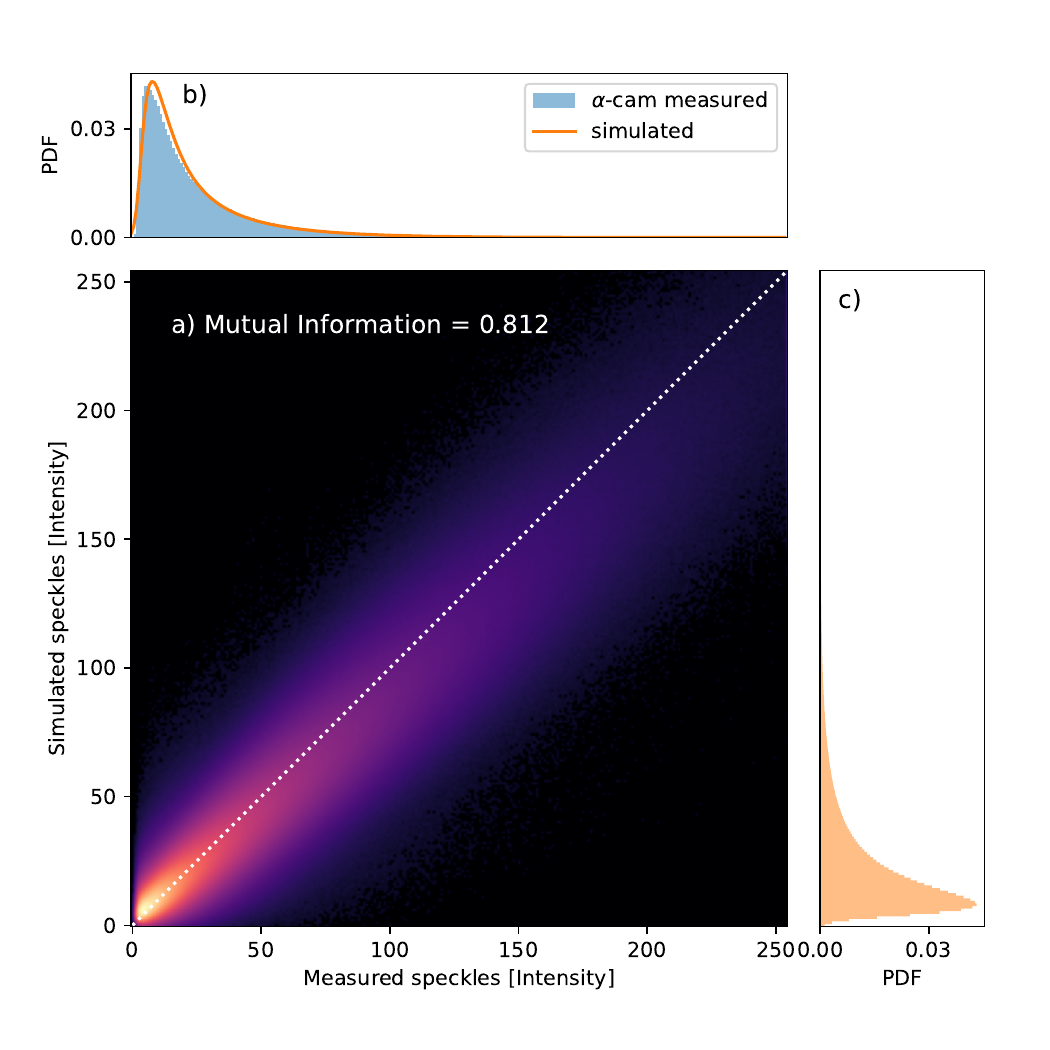}
    \includegraphics[width=1.\linewidth]{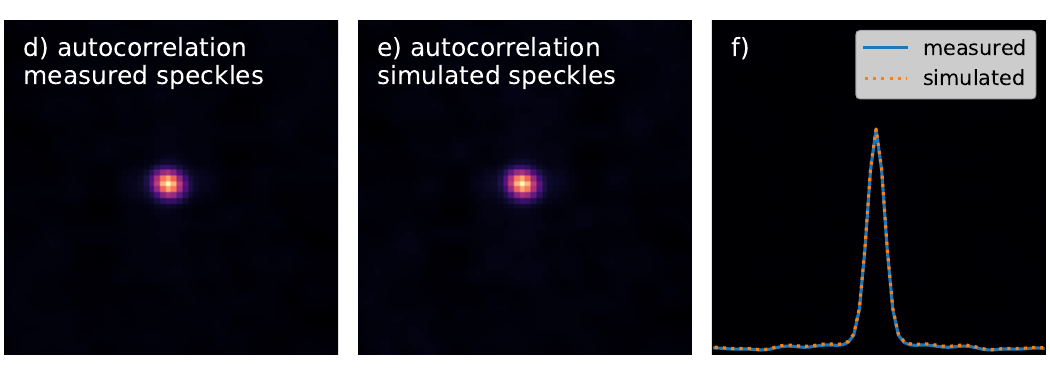}
    \caption{
    a) Bi-dimensional intensity histogram between measured and simulated speckles. The diagonal is the ideal histogram map when simulation perfectly matches the measured data. One can notice that dispersion occurs, instead, quantified by a mutual information $0.812$, cf. Eq. (\ref{def:mi}).
    b) Histogram plot of the measured speckle intensities, projection along the vertical axis of the 2D histogram. In orange, we superimpose the plot of the histogram of the simulated speckles. 
    c) Intensity histogram of the simulated speckles, 2D histogram projection along horizontal axis.
    d) Average autocorrelation of the measured speckles versus 
    e) autocorrelation of the simulated speckle pattern using the inferred $\mathbb{T}$.
    f) Autocorrelation difference (dark image in the background) and central profile plot of the two functions proving practically identical average speckle size recovered after the transmission characterization.
    }
    \label{fig:histo2D}
\end{figure}

\paragraph*{Real fiber propagation versus TM simulation.}
The MMF is typically treated as a linear complex random projector, and its transmission rule could be estimated by finding the transmission matrix. 
In the case of a good $\mathbb{T}$ recovery, one would expect that the speckles simulated given a certain input closely match  the experimentally recorded output by the  camera. 
Consequently, training a classifier with the simulated output should give performance that are similar to those obtained with the real data. 
However, Fig.~\ref{fig:accuracy}b highlights a strong discrepancy in accuracy with the simulations and Fig. \ref{fig:local_energy} suggest a different local energy profile.
This is a surprising fact that is worth investigating further.
For this qualitative analysis, we use the data from experiment 2 which was specifically designed to recover the transmission matrix.

In Fig. \ref{fig:speckleout}a, we show a representative output speckle pattern recorded by $\alpha$-cam. 
For better clarity, we restrict our analysis to a portion of the whole speckle output, identified with a red box and shown in Fig. \ref{fig:speckleout}b.
The result of the simulation is reported Fig.\ref{fig:speckleout}c and display the reconstructed speckle pattern originated from a random input patter which was included in the training set.
Another representative pattern, not included in the training is shown in Fig. \ref{fig:speckleout}e, together with its corresponding simulated version Fig. \ref{fig:speckleout}f.
For both, we observe minimal discrepancies between real and simulated data, which we quantify by plotting the difference map between the two (Fig. \ref{fig:speckleout}d,g).
As an additional check, we also compute the focusing operator $\mathbb{T}\mathbb{T}^\dagger$, that we report in Fig.\ref{fig:speckleout}h. 
The diagonality of the norm of this operator is normally used for testing the fidelity of the recovered transmission matrix \cite{popoff2010measuring}. 
\textcolor{black}{
In Fig. \ref{fig:histo2D}, we also compared the distribution of measured and simulated speckle intensities computing the 2D histogram distribution (panel \ref{fig:histo2D}a) and its relative marginalizations (the histograms of the intensity distributions for each dataset, which is the integral of the 2D histogram along the two directions, Fig. \ref{fig:histo2D}b,c).
For completeness, since the 2D histogram is normally used to calculate the mutual information between the two datasets, we also report its value. 
Additionally, we analysed the average autocorrelation of the speckles both from the measured data and the synthetic data created using the inferred transmission matrix (Fig. \ref{fig:histo2D}d,e,f).
}
From the histogram analysis, a perfect match between measured and simulated data would have produced a 2D-histogram map with only diagonal entries. 
\textcolor{black}{
The fact that the diagonal is broadened implies that the correspondence between the measured intensities and the simulated dataset is not entirely captured by the recovery of the linear transmission, even if the speckles are effectively reproduced (as in Fig. \ref{fig:speckleout}).
}

To further restrict the reason for this discrepancy, we analyzed the average speckle autocorrelation of the measured (Fig. \ref{fig:histo2D}d) versus simulated (Fig. \ref{fig:histo2D}e) dataset. We notice that the overall autocorrelation shape is very similar, and the profile plot in panel f confirms the close matching between the datasets.
Since the autocorrelation is directly connected with the average size of the coherence region of a single speckle grain, having the same autocorrelation implies the same statistical spatial distribution of the two speckle patterns, which then could accommodate a comparable number of optical modes.
\textcolor{black}{As an additional test, we decided to simulate the speckle output using a random-phase (flat distribution $\in \left[0-2\pi \right)$) complex-valued transmission matrix (keeping the modulus as retrieved in the experiment) and test its classification performance. 
This new dataset performs similarly to the randomized MNIST (see supplementary code in the online repository, App. \ref{subsec:code}), not reaching the experimental results.
}

\begin{figure}
    \centering
    \includegraphics[width=1.\linewidth]{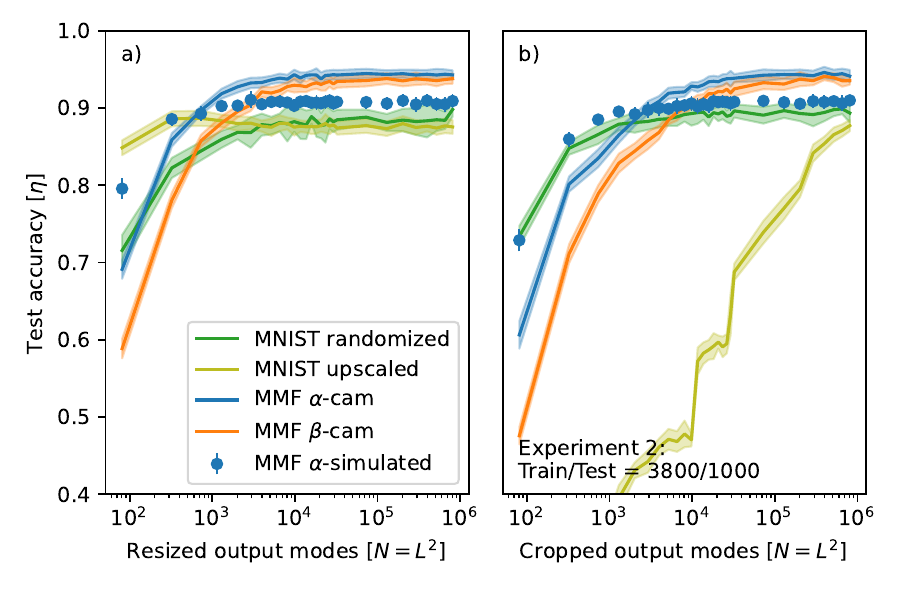}
    \caption{
    Training accuracy trade-off when reducing the number of output fiber modes in experiment 2.
    We study the performances obtained using de-magnified camera measurements as a function of their output size (blue and orange plots).
    With dots, we report the same study performed with simulated speckle patterns. In green and olive, respectively, the results for the upscaled and randomized MNIST datasets. 
    In a), the study is done by resizing the output patterns, and 
    \textcolor{black}{in b) a similar study is done by cropping windows smaller than the original dimension down to different sizes $L$ (thus excluding peripherical speckles). Along the x axis we report the number of the maximum optical modes allowed after resizing and cropping, $N=L^2$.}
    For both, we notice that performances remain stable down to a substantial reduction of the number of modes used in the training set (around 400 pixels, 80\% less pixel than the full resolution dataset). 
    }
    \label{fig:zoomcrop}
\end{figure}

\paragraph*{Influence of the number of modes.}
As a last analysis, we evaluate the effect of the number of output modes in two different ways.
In Fig.~\ref{fig:zoomcrop}a, we evaluate the effect of downscaling the MMF-output of experiment 2 and, in panel b, cropping it to a smaller window of increasing size.
The effect of these operation is that we variate the size $L$ of the output dataset used to train the classificator and, accordingly, the total number of the output modes $N=L^2$.
For both camera detections, reducing the number of modes has a negative impact on the performance with the cropping operation being more drastic than rescale.
At around $L=400$ pixels, however, both operations had similar effect with performance nearly identical to the full resolution image but with reduced numerical complexity.
The fact that the output downscaled by a factor of around $2$ has similar performance to the full resolution dataset seems in agreement with the fact that the spatial correlation of the speckle patter is wider than a single pixel in the detected image, thus introducing redundant information that can be compressed.
We report, however, that this also happens with the cropped version of the output, which still shares the same spatial properties of the average speckle size.
Remarkably, we also register that the fiber simulation does not perform equally well, with the only exception at very small sizes (up to L=36) when accuracy is still low and of no practical usage.
Furthermore, we notice that the other datasets (randomized and upscaled) still perform worse compared to the hardware fiber after $L=54$, even if at these regime the accuracy obtained is relatively low. 
Additionally, from Fig. \ref{fig:zoomcrop}a, we observe that upscaling the original MNIST has a negative impact on the performance possibly due to overfitting, being the ideal dimension of the dataset sitting at around $L=18-32$ (local maximum of the curve). 
This explains also the lower performance registered in Fig. \ref{fig:accuracy}.
On the other hand, in Fig. \ref{fig:zoomcrop}b, the same dataset has a dramatic dependence on cropping. This can be expected, because with the crop we are restricting the observation window down to a small feature of the number-image, not capturing its entire shape. 
Among these options, we can operatively conclude that best way to improve classification accuracy is by using a hardware MMF projector.

\section{Perspectives}
\label{sec:Perspectives}
In this work, we used MMFs to realize random transformations of the MNIST dataset showing that a linear classifier has better accuracy on the MMF-transformed dataset than on the original one. 
Complementary to high-intensity pulsed excitation \cite{teugin2021scalable}, this transformation (MMF and camera detection) is non-linear even in continuous low-power regime and increases the dimension of the data, but those characteristics are not enough to justify improved accuracy alone.
In fact, data upscaling (that increases the dimension), random matrix multiplication (that projects on random space), and the MMF simulation did not reach performances similar to the transformation provided by the physical MMF.
As anticipated in section \ref{sec:Results}, our goal was not to compete with the accuracy of more sophisticated architectures, but rather showing that MMFs are simple —yet robust— hardware solutions for optical computing.
For example, convolutional neural networks exploit spatial correlations in the data and work particularly well for image datasets.
Our approach is, instead, closer to that of a fixed-weight densely-connected network, leaving room for applicability to different variety of data types.
However, contrarily to general random transformations (that destroy spatial correlations), the fiber output presents a correlation property determined by the average size of the speckle patterns.

An MMF used as a physical neural network is cheap, can flexibly be mounted to deliver light in a user-defined position, and offers a different set of random projections each time it is re-positioned (thus requiring independent training of the output layer).
Indeed, we still need an SLM and at least one camera to record the speckled projection, but those are almost unavoidable in any ONNs configuration. 
To the best of our knowledge, the current state-of-the-art is achieved using FPGA hardware in conjunction with data augmentation, reaching 98\% test accuracy \cite{moran2020}.
Another approach using disordered optical media exploits polaritons to reach 96\% accuracy \cite{mirek2021neuromorphic}, which is comparable with average optimizations obtained with the MMF approach.
These results strengthen the fact that MMFs are promising tools for neuromorphic computing, with the additional advantage of their simplicity and easiness of use.
Further, we believe that our results could be relevant for the theoretical understanding of deep neural networks: in the spirit of random-feature models \cite{gerace2020generalisation,goldt2022gaussian,baldassi2021learning}, we showed that the class in which we sample the random features plays important role in the accuracy, as suggested in \cite{perugini2022}. 
In fact, while taking a Gaussian random matrix already improves the accuracy a bit, the transformation implemented by an MMF makes a much bigger difference.
Further investigation is needed to understand why the specific hardware transformation provided by the MMF is so effective. 
In particular, the local energy profiles suggest that this effectiveness could be explained by studying wide flat regions in the loss landscape, in the same spirit as \cite{baldassi2021learning}: 
to do so, the authors use a quantity called \emph{local entropy}, which is only approximated by the local energy that we discussed here (this might explain the discrepancy between the local energy profile of the upscaled dataset in \ref{fig:local_energy} and its test error in \ref{fig:accuracy}).
Here, we put forward some conjectures based on the present study. 
First, the fact that the accuracy gap between the physical MMF data and its simulation (Fig.~\ref{fig:accuracy}) is reflected in the local energy profile (Fig.~\ref{fig:local_energy}) makes us confident that the two approaches indeed belong to different classes of random transformations. 
The fact the physical MMF transformation is so robust to perturbations is consistent with the great redundancy of the data that emerges from Fig.~\ref{fig:zoomcrop} and Fig.~\ref{fig:histo}, where we see that we can delete the majority of the signal before losing accuracy. 
We conjecture that the random transformation realized by MMFs leads to well-separated projections in the high-dimensional space that allow for a good classification accuracy that is also resistant to noise, in a way that is reminiscent of error-correcting codes.
All these considerations highlight the need to further investigate how those widespread MMF devices can be modeled and exploited in particular in the design of optical neural networks.

\begin{acknowledgments}
We thank Dr. Raffaele Marino for useful discussions.
We acknowledge the support of LazioInnova - Regione Lazio under the program {\em Gruppi di ricerca 2020} - POR FESR Lazio 2014-2020, Project NanoProbe (Application code A0375-2020-36761), the support from the European Research Council (ERC) under the European Union’s Horizon 2020 Research and Innovation Program, Project LoTGlasSy (Grant Agreement No. 694925, Prof. Giorgio Parisi) and the support 
of the PRIN project ``Complexity, disorder and fluctuations'' of the Italian Ministry of University and Research (MUR) under the NextGenerationEU program, code No. 2022LMHTET.
\end{acknowledgments}

\appendix
\subsection{Neural network architecture and training procedure}
\label{app:nn_details}
Our classification model involves a potentially fully-connected  layer (in the sense that we do not restrict any intensity mode couplings) that maps linearly the $28 \times 28$ image space into a higher dimensional $N=900 \times 900$ output space. 
On the output space, we build a linear classification model using the \textit{LogisticRegression} function provided by the python library RapidsAi \cite{rapids}, the GPU equivalent of the Scikit-Learn implementation. 
Given an input pattern $\{\xi_i\}_{i=1,..,N}$ the \textit{LogisticRegression} function performs a weighted average of the $N$ input channels, producing a score $z_j=\sum_{i=1}^N w_{ji} \xi_i $ for each of the $10$ classes corresponding to each type of digit. 
The scores are then transformed to probabilities with
\begin{equation}
    p_j=\frac{e^{z_j}}{\sum_{j'=1}^{10} e^{z_{j'}}}
\end{equation}
and plugged in a cross-entropy loss function that is a sum of contributions coming from all of the $P$ input patterns that we are using to train the model
\begin{equation}
    L(\mathbf{w})= - \sum_{\mu=1}^P \log (p_{j^*})
\end{equation} 
where $j^*$ is the index of the correct class of each input pattern. 
The cross-entropy loss $L(\mathbf{w})$ is then minimized with a gradient-descent-related strategy to find the configuration of the weights $\mathbf{w}^*$ that has the highest classification accuracy.
\textcolor{black}{
The classification accuracy is defined as the fraction of correctly classified entries divided by the total number of training (or test) images. 
To compute it after the parameter optimization, we make use of the \texttt{sklearn.metrics.accuracy\_score} function of the Scikit-learn library.
}

\subsection{Measure of local energy}
\label{app:local_energy_appendix}

Given a loss (namely energy) function $L$ that depends on a set of parameters $\mathbf{w}$, we define the local energy as the following expectation value
\begin{equation}
    L_{\texttt{local}}(\sigma)=\mathbb{E}_{\eta_{ij} \sim \mathcal{N}(0,\sigma)} L(\{w_{ij}\eta_{ij}\})
\end{equation}
where $\{\eta_{ij}\}$ is a set of i.i.d. random Gaussian variables with zero mean and variance $\sigma$ that multiply element-wise the set model parameters $\{w_{ij}\}$. The local energy $L_{\texttt{local}}(\sigma)$ still depends on the variance $\sigma$ of the Gaussian noise. As explained in the main text, we are interested in studying how quickly the local energy increases when we increase $\sigma$: from the literature (see main text) we know that a slower increase is correlated with a higher test accuracy. 
\textcolor{black}{For Fig.~\ref{fig:local_energy} of the main text we choose $L$ as the fraction of misclassified data point in the train set.}

\subsection{Experimental setup}
The sketch of the experimental setup is shown in Fig. \ref{fig:scheme}c.
In the experiment, we used a continuous Melles Griot he-ne laser ($632.8 nm$) as light source. 
The emitted beam is magnified 15 times through a $5:75 cm$ telescope before being imprinted on a Hamamatsu SLM in polarization configuration (model LCOS-SLM x10488 series, pixel size: $20 \mu m$).
The real space plane of the SLM is then recreated on the entrance facet of the optical fiber using a pair of $50:7.5 cm$ focal lenses after the spatially modulated beam profile has been collected.
Thorlabs FT-1.0-EMT, $NA=0.39$, $1$ meter long, 1mm core multi-mode optical fiber is the one that has been utilized.
We indicate each facet of the MMF with the letter $\alpha$ and $\beta$.
Two IDS cameras (UI-5370CP-M-GL and UI-5480CP-M-GL) with pixel sizes of $5.5$ and $2.2 \mu m$, respectively, are used to collect the counter-polarized (respect to the laser) reflection from the injection surface as well as the transmission signal. 
To achieve the same spatial resolution of $1.1 \mu m / px$ on both cameras, the magnification was set to $5\times$ and $3\times$ respectively. 
\textcolor{black}{
The MNIST handwritten digits and random masks are sent to the SLM in alternated sequences and are encoded in the same way. 
In practice, for each of those, we send an image (random or MNIST) having a size of $28 \times 28$ pixels, focusing it so that it is inscribed on the input facet of the optical fiber.
Each pixel uses grayscale values ranging from $0$ to $10$.
}
The random patterns are sent for the sole purpose of the characterization of the fiber transmission, and are not used for training of the classification layer.
The light propagating through this disordered optical device reaches both edges and produces a seemingly random interference pattern of intensities (the speckles).

\subsection{Number of optical modes.}
\textcolor{black}{
The fiber used (FT1000EMT, Thorlabs) has a diameter of $d=1 mm$ with a $NA=0.39$. Thus, the maximum theoretical number of supported modes is $N_{modes} =  (\pi d NA / \lambda)^2 /2$, which gives around $1.871 \cdot 10^3$ modes.
For the experimental realization, the number of optical modes is influenced by the number of camera pixels used to record the fiber’s output and the average physical size of the speckles. 
In our case, the average full-width half maximum of the speckles is 1px, and using a squared portion of the central core of the fiber having $L=600$ determines a maximum total number of imaged modes equal to $L^2= 360 \cdot 10^3$ modes.
This is a reduced fraction of the total number of imaged modes of the entire facet, consisting of about $635 \cdot 10^3$ modes.
}

\subsection{Under-exposure, camera saturation, and measurement stability}
\label{app:underexposure}
When setting the exposure time of the camera, we are implicitly acting on the way it records the signal. 
If the exposure time is fast enough with respect to the intensity delivered, the camera underexposes the signal, i.e. does not detect the signal in a particular region.
The opposite effect, over-exposure, happens when intesity is too high compared to a long exposure of the images.
In both cases, a non-linear threshold is introduced in the detected signals.
To try to assess its effect on the classifier accuracy, we tried to explore several intensity distributions of the datasets recorded in camera. 

\begin{figure}[t!]
    \centering
    \includegraphics[width=1.\linewidth]{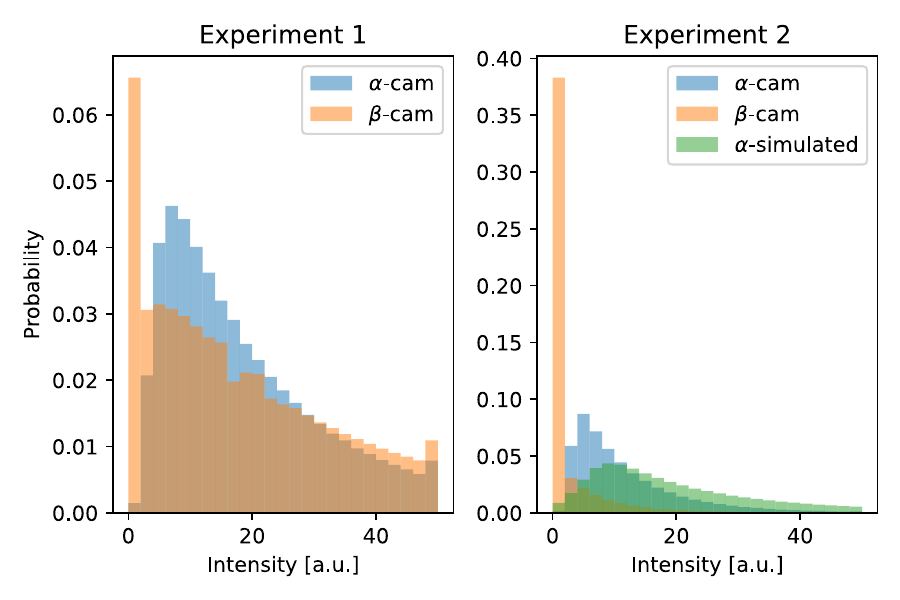}
    \caption{
    Intensity distribution of the speckle pattern measured in camera $\alpha$ and $\beta$ during two independent experiments. 
    Given a stable laser output, we modify the camera exposure time to force a certain amount of non-linearity (in the form of a recording threshold) in the measuring process.
    a) In experiment 1, the dynamic range of the camera $\alpha$ fits well the intensity distribution of the speckle images recorded, whereas camera $\beta$ underexposes around $7\%$ of the signal.
    b) In experiment 2, the camera $\alpha$ have a similar trend to that of experiment 1, but in camera $\beta$ we strongly underexpose the images, cutting out $37\%$ of the light intensity reaching the sensor.}
    \label{fig:histo}
\end{figure}

In experiment 1, Fig. \ref{fig:histo}a, $\alpha$-cam provides an optimal dynamic range, with low under- (0.1\%) and over-exposure (1\%). 
Instead, the $\beta$-cam recorded the signal underexposing 7\% of the total pixels in the image.
In experiment 2, Fig. \ref{fig:histo}b, the $\alpha$-cam correctly sample the intensities, whereas $\beta$-cam is set to cut off 37\% of the pixels. 
Additionally, we report also the intensity distribution obtained with the simulation of the light propagating through the fiber and detected by the $\alpha$-cam.
We notice a substantial difference between the intensity distribution of the recorded and simulated data: this could explain the different performances achieved by the two datasets.

\begin{figure}
    \centering
    \includegraphics[width=1.\linewidth]{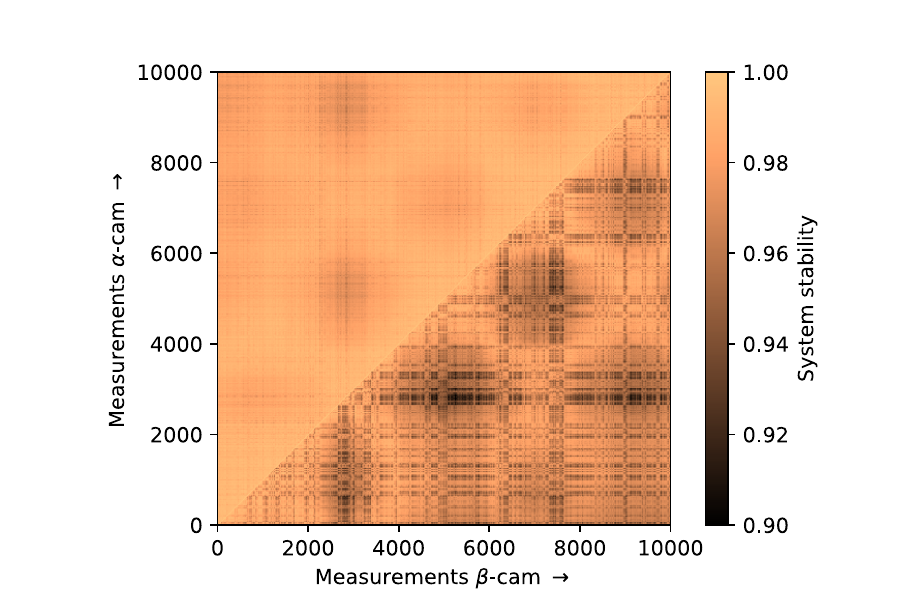}
    \caption{Measurement stability during experiment 1. 
    Keeping the initial probing frame, we compute the normalized scalar product against the output of the same frame at different time for both $\alpha,\beta$ cams.
    The upper half of the plot is the correlation stability of the $\alpha$-cam (thus how similar the output look when the same input is sent again during the experiment) and the bottom one for $\beta$.
    In both cases the correlation remained higher than $90\%$.
    }
    \label{fig:overlap}
\end{figure}

Over the entire duration of the experiment, we continuously monitored the fiber stability by sending to the SLM an identical image.
When the fiber is sufficiently stable, the speckle pattern produced at the facets must be always identical to the ones recorded at the beginning of the experiment.
Keeping the camera frame $\tau=0$ as a reference for both $\alpha,\beta$ cameras, we computed the normalized scalar product against the speckle image at a given time $\tau'$:
\begin{equation}
    \rho\left(\tau,\tau'\right) = \frac{s_\tau^{\{\alpha,\beta\}} \cdot s_{\tau'}^{\{\alpha,\beta\}}}{{\left|s_\tau^{\{\alpha,\beta\}}\right|}{\left|s_{\tau'}^{\{\alpha,\beta\}}\right|}},
\label{eq:overlap}
\end{equation}
where we called $s$ the speckle pattern recorded at the time.
Using this metric, $\rho \approx 1$ means the measurements are highly correlated whereas $\rho \approx 0$ implies that the system decorrelated during the measurement.
Fig. \ref{fig:overlap} reports the stability study across the entire duration of the experiment 1. 
We notice that the $\alpha$-cam remains highly correlated ($96\%$ minimum), compared to the $\beta$-cam (90\% minimum).
Despite the lower correlation stability and $7\%$ underexposed pixels values, the $\beta$-cam resulted as accurate as the $\alpha$ cam during the classification of the test set (Fig.\ref{fig:accuracy}).

In Fig. \ref{fig:histo2D} we compare the output speckles corresponding to the same input through the real MMF and through a synthetic MMF whose transmission matrix is the one inferred from data by phase retrieval. The two do not appear to be the same, that is their scatter plot is not exactly diagonal. We quantify their mutual difference by means of the mutual information
\begin{equation} \label{def:mi}
   I({\rm real}|{\rm synth})\equiv  \sum_{i=1}^{256} P_{\rm real}(y_i) \log \frac{P_{\rm real}(y_i)}{P_{\rm synth}(y_i)}, 
\end{equation}
where $256$ are the intensity bins. A perfect match would yield $I=1$, whereas in Fig. \ref{fig:histo2D}a we find $I=0.812$.

\subsection{Code and dataset availability}
\label{subsec:code}
\textcolor{black}{
The code to reproduce the results on Fig.\ref{fig:accuracy} is freely downloadable from the github at \href{https://github.com/danieleancora/MMFclassification.git}{danieleancora/MMFclassification.git}
and the relative datasets are available from FigShare with the DOI:
\href{https://doi.org/10.6084/m9.figshare.25551186.v1}{10.6084/m9.figshare.25551186.v1}. 
}

\bibliography{apssamp}
\end{document}